\documentclass[prb,twocolumn,showpacs,preprintnumbers,amsmath,amssymb]{revtex4}

\usepackage{graphicx}
\usepackage{dcolumn}
\usepackage{bm}
\usepackage{color}

\newcommand{\mapright}[1]{\smash{\mathop{\hbox to 1.0cm{\rightarrowfill}}\limits^{#1}}}

\begin{document}

\preprint{}

\title{ Theory of quantum transport in Josephson junctions with a ferromagnetic insulator}

\author{Shiro Kawabata$^{1,2}$ and Yasuhiro Asano$^{3}$}

\affiliation{
$^1$Nanosystem Research Institute (NRI), National Institute of Advanced Industrial Science and Technology (AIST), Tsukuba, Ibaraki, 305-8568, Japan
\\
$^2$CREST, Japan Science and Technology Corporation (JST), Kawaguchi, Saitama 332-0012, Japan 
\\
$^3$Department of Applied Physics, Hokkaido University,
Sapporo, 060-8628, Japan
}

\date{\today}

\begin{abstract}

We investigate the Josephson transport through ferromagnetic insulators  (FIs) by taking into account the band structure of FIs explicitly.
Using the recursive Green's function method, we found the formation of a $\pi$-junction in such systems.
Moreover the atomic-scale 0-$\pi$ oscillation is induced  by increasing the thickness of FI and its oscillation period  is universal, i.e., just single atomic layer.
Based on these results, we show that stable $\pi$-state can be realized in junctions
based on high-$T_c$ superconductors with La$_2$BaCuO$_5$ barrier.
Such FI-based Josephson junctions may become an element in the architecture of future quantum computers.
\end{abstract}

\maketitle

\section{Introduction}
There is an increasing interest in the novel properties of interfaces and junctions of superconductors and ferromagnetic materials.\cite{rf:Golubov,rf:Buzdin1}
One of the most interesting effects is the formation of a  Josephson $\pi$-junction in superconductor/ferromagnetic-metal/superconductor (S/FM/S) heterostructures.\cite{rf:Buzdin2}
In the ground-state phase difference between two coupled superconductors is $\pi$ instead of 0 as in the ordinary 0-junctions.
In terms of the Josephson relationship 
\begin{eqnarray}
I_J= I_C \sin \phi,
\end{eqnarray}
 where $\phi$ is the phase difference between the two superconductor layers, a transition from the 0 to $\pi$ states
implies a change in sign of $I_C$ from positive to negative. 
Such a negative $I_C$  was originally found in the Josephson effect with  a spin-flip process.\cite{rf:Kulik,rf:Shiba,rf:Bulaevskii} 
In S/FM/S junctions, such a sign change of $I_C$  is a consequence of a phase change in the pairing wave-function induced in the FM layer due to the proximity effect.
 The existence of the $\pi$-junction in S/FM/S systems has been confirmed in experiment by Ryanzanov et al.\cite{rf:Ryanzanov} and Kontos et al.\cite{rf:Kontos}

Recently, a quiet qubit consisting of a superconducting loop with a S/FM/S $\pi$-junction has been proposed.\cite{rf:Ioffe,rf:Blatter,rf:Yamashita}
In the quiet qubit, a quantum two-level system (qubit) is spontaneously generated and therefore it is expected to be robust to the decoherence by the fluctuation of the external magnetic field.
From the viewpoint of the quantum dissipation, however, the structure of S/FM/S junctions is inherently identical with S/N/S junctions (N is a normal nonmagnetic metal).
Thus a gapless quasiparticle excitation in the FM layer is inevitable.
This feature gives a strong dissipative effect\cite{rf:Zaikin,rf:Schon,rf:Kato} and the coherence time of S/FM/S quiet qubits is bound to be very short.
Therefore Josephson $\pi$ junctions with a nonmetallic interlayers are highly desired for qubit application.

On the other hand, a possibility of the $\pi$-junction formation in Josephson junctions through ferromagnetic insulators (FIs) have been theoretically predicted\cite{rf:Tanaka} and intensively analyzed by use of the quasiclassical Green's function techniques.\cite{rf:Fogelstrom,rf:Lofwander}
Recently, by extending these results, we have proposed  superconducting phase\cite{rf:Kawabata1} and flux qubits\cite{rf:Kawabata2,rf:Kawabata3,rf:Kawabata4} based on S/FI/S $\pi$-junctions.
 Moreover we have also showed that the effect of the dissipation due to a quasi-particle excitation on macroscopic quantum tunneling is negligibly small.\cite{rf:Kawabata3}
These results clearly indicate the advantage of the FI based $\pi$-junction  for qubit applications with longer coherence time.

However, up to now, a simple $\delta$-function potential\cite{rf:Tanaka} has been used in order to model the FI barrier.
In this phenomenological model, the up (down) spin electrons tunnel through a positive (negative) delta-function barrier.
Therefore, strictly speaking, this model describes not  ferromagnetic insulators but half metals with infinitesimal thickness.
Moreover the possibility of the $\pi$-junction formation in the finite barrier thickness case  is also an unresolved problem.
In order to resolve above issues, we formulate a numerical calculation method for the Josephson current through FIs by taking into account the band structure and the finite thickness of FIs explicitly.
In this paper we present our recent numerical results\cite{rf:Kawabata4,rf:Kawabata5,rf:Kawabata6} on the formation of the $\pi$-coupling for the Josephson junction through a FIs, e.g., La${}_2$BaCuO${}_5$ and K${}_2$CuF${}_4$ and show that the mechanism of the $\pi$-junction in such systems is in striking contrast to  the  conventional S/FM/S junctions.

\section{Magnetic and Electronic properties of ferromagnetic insulators}

In this section, we briefly describe the magnetic properties and the electronic density of states (DOS) of FIs.
The typical DOS of FI for each spin direction is shown schematically in Fig. 1.
One of the representative material of FI is half-filled La${}_2$BaCuO${}_5$ (LBCO).\cite{rf:Mizuno,rf:Masuda,rf:Ku}
The crystal structure of LBCO has tetragonal symmetry with space group $P4/mbm$.
In 1990, Mizuno $et$ $al$, found that LBCO undergoes a ferromagnetic transition at 5.2 K.\cite{rf:Mizuno}
The exchange splitting $V_\mathrm{ex}$ is estimated  to be 0.34 eV by a first-principle band calculation using the spin-polarized local density approximation.\cite{rf:LBCO}
Since the exchange splitting is large and the bands are originally half-filled, the system becomes FI. 

\begin{figure}[t]
\begin{center}
\includegraphics[width=7.5cm]{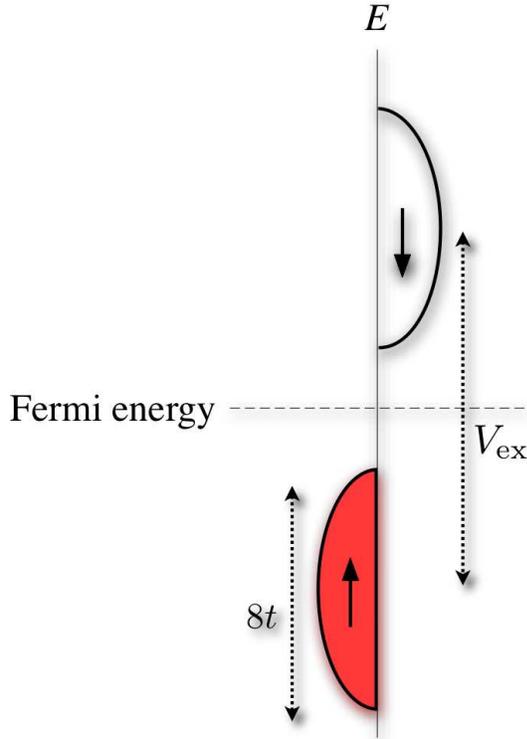}
\end{center}
\caption{The density of states for each spin direction for a ferromagnetic insulator, e.g., LBCO. $V_\mathrm{ex}$ is the exchange splliting and $8 t$ is the width of the band.
 }
\label{fig1}
\end{figure}

An another example of FPFI is K${}_2$CuF${}_4$ compounds in which the two-dimensional Heisenberg ferromagnet is realized.\cite{rf:Moreira1,rf:Moreira2}
The ferromagnetic behavior of this materials has been experimentally confirmed by the magnetic susceptibility\cite{rf:Yamada} and neutron diffraction measurements.\cite{rf:Hirakawa}
Moreover a result of the first-principle band calculation\cite{rf:Eyert} indicated that K${}_2$CuF${}_4$ compounds with Jahn-Teller distortion have the electronic structure similar to Fig. 1. 
In the followings, we calculate the Josephson current through such FIs numerically.

\section{Numerical method}
\begin{figure}[tb]
\begin{center}
\includegraphics[width=8.5cm]{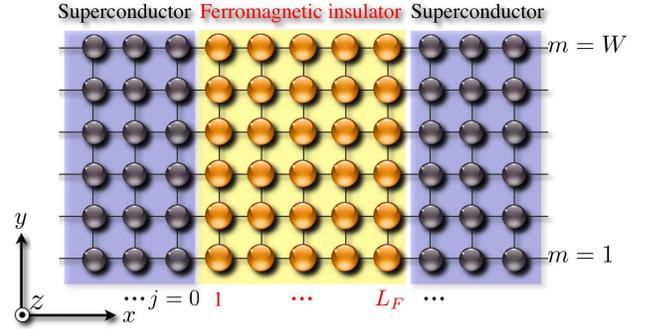}
\end{center}
\caption{A schematic figure of a Josephson junction through the ferromagnetic insulators on the
two-dimensional tight-binding lattice. 
 }
\label{fig2}
\end{figure}

In this section, we develop a numerical calculation method for the Josephson current of S/FI/S junctions based on the recursive Green's function technique.\cite{rf:Furusaki,rf:Asano1,rf:Asano2}
Let us consider a two-dimensional tight-binding model for the S/FI/S junction as shown in Fig.~2.
The vector 
 \begin{eqnarray}
 \boldsymbol{r}=j{\boldsymbol{x}}
+m{\boldsymbol{y}}
\end{eqnarray}
 points to a lattice site, where ${\boldsymbol{x}}$ and ${\boldsymbol{y}}$ are unit vectors in the $x$ and $y$ directions,
respectively.
In the $y$ direction, we apply the periodic boundary condition for the number of lattice sites being $W$.

Electronic states in a superconductor are described by the
mean-field Hamiltonian
 \begin{eqnarray}
{\cal H}_{\text{BCS}}&=& \frac{1}{2}\sum_{\boldsymbol{r},\boldsymbol{r}^{\prime }  \in \text{S}    }%
\left[ \tilde{c}_{\boldsymbol{r}}^{\dagger }\;\hat{h}_{\boldsymbol{r},%
\boldsymbol{r}^{\prime }}\;\tilde{c}_{\boldsymbol{r}^{\prime }}^{{}}-%
\overline{\tilde{c}_{\boldsymbol{r}}}\;\hat{h}_{\boldsymbol{r},\boldsymbol{r}%
^{\prime }}^{\ast }\;\overline{\tilde{c}_{\boldsymbol{r}^{\prime }}^{\dagger
}}\;\right]
\nonumber\\
 &+&
 \frac{1}{2}\sum_{\boldsymbol{r}\in \text{S}}\left[ \tilde{c}_{%
\boldsymbol{r}}^{\dagger }\;\hat{\Delta}\;\overline{\tilde{c}_{\boldsymbol{r}%
}^{\dagger }}-\overline{\tilde{c}_{\boldsymbol{r}}}\;\hat{\Delta}^{\ast }\;%
\tilde{c}_{\boldsymbol{r}}\right]   
.
\end{eqnarray}
Here
 \begin{eqnarray}
\hat{h}_{\boldsymbol{r},\boldsymbol{r}^{\prime }}&=& \left[ -t_s \delta _{|%
\boldsymbol{r}-\boldsymbol{r}^{\prime }|,1}+(-\mu_s
+4t_s)\delta _{\boldsymbol{r},\boldsymbol{r}^{\prime }}\right] \hat{\sigma}_{0}
,
\end{eqnarray}
with 
 \begin{eqnarray}
\overline{\tilde{c}}_{\boldsymbol{r}}=\left( c_{\boldsymbol{r}%
,\uparrow },c_{\boldsymbol{r},\downarrow }\right) ,
\end{eqnarray}
 where
  $
  c_{\boldsymbol{r} ,\sigma }^{\dagger }$ ($c_{\boldsymbol{r},\sigma }^{{}}
$)
 is the creation
(annihilation) operator of an electron at $\boldsymbol{r}$ with spin $\sigma
=$ ( $\uparrow $ or $\downarrow $ ), $\overline{\tilde{c}}$ means the
transpose of $\tilde{c}$,  and $\hat{\sigma}_{0}$ is $2\times 2$ unit matrix. 
The chemical potential  $\mu_s$ is set to be $2 t_s$ for superconductors.
In superconductors, the hopping integral $t_s$ is considered among nearest neighbor sites and we choose 
 \begin{eqnarray}
\hat{\Delta}=i\Delta \hat{\sigma}_{2}
,
\end{eqnarray}
where $\Delta $ is the amplitude 
of the pair potential in the $s$-wave or $d$-wave symmetry channel, and $\hat{\sigma}_{2}$ is a Pauli matrix.

We consider FIs as a barrier of the Josephson junction.
The Hamiltonian of the FI barrier is given by a single-band tight-binding model as 
 \begin{eqnarray}
{\cal H}_\mathrm{FI} &=& -t \sum_{\boldsymbol{r},\boldsymbol{r}^{\prime },\sigma} 
c_{\boldsymbol{r},\sigma}^\dagger 
c_{\boldsymbol{r}',\sigma}
-\sum_{\boldsymbol{r}} ( 4 t -\mu)  
c_{\boldsymbol{r},\uparrow}^\dagger 
c_{\boldsymbol{r},\uparrow}
\nonumber\\
&+&
 \sum_{\boldsymbol{r}} 
( 4 t -\mu + V_\mathrm{ex}) 
 c_{\boldsymbol{r},\downarrow}^\dagger 
 c_{\boldsymbol{r},\downarrow}
,
  \end{eqnarray}
where $V_\mathrm{ex}$ is the exchange splitting (see Fig. 1).
If $V_\mathrm{ex} >  8 t$ ($V_\mathrm{ex} <  8 t$), this Hamiltonian describes FI (FM).
The chemical potential $\mu$ is set to be 
 \begin{eqnarray}
\mu=\frac{V_\mathrm{ex}}{2}  + 4t
.
  \end{eqnarray}

The Hamiltonian is diagonalized by the Bogoliubov transformation and the
Bogoliubov-de Gennes equation is numerically solved by the recursive
Green function method.\cite{rf:Furusaki,rf:Asano1,rf:Asano2} 
We calculate the Matsubara
Green function in a FI region,
\begin{equation}
\check{G}_{\omega _{n}}(\boldsymbol{r},\boldsymbol{r}^{\prime })=\left(
\begin{array}{cc}
\hat{g}_{\omega _{n}}(\boldsymbol{r},\boldsymbol{r}^{\prime }) & \hat{f}%
_{\omega _{n}}(\boldsymbol{r},\boldsymbol{r}^{\prime }) \\
-\hat{f}_{\omega _{n}}^{\ast }(\boldsymbol{r},\boldsymbol{r}^{\prime }) & -%
\hat{g}_{\omega _{n}}^{\ast }(\boldsymbol{r},\boldsymbol{r}^{\prime })%
\end{array}
\right) , \label{deff}
\end{equation}
where 
\begin{equation}
\omega _{n}=(2n+1)\pi T
\end{equation}
 is the Matsubara frequency, $n$ is an
integer number, and $T$ is a temperature. 
The Josephson current is given by
\begin{equation}
I_J (\phi)=-ietT\sum_{\omega _{n}}\sum_{m=1}^{W}\mathrm{Tr}\left[ \check{G}_{\omega
_{n}}(\boldsymbol{r}^{\prime },\boldsymbol{r})-\check{G}_{\omega _{n}}(%
\boldsymbol{r},\boldsymbol{r}^{\prime })\right]
,
\end{equation}
with $
\boldsymbol{r}^{\prime }=\boldsymbol{r}+\boldsymbol{x}$.
The Matsubara Green function in Eq.~(\ref{deff}) 
is a $4\times 4$ matrix representing Nambu and spin spaces. 
Throughout this paper we fix $T=0.01T_{c}$, where $T_c$ is the superconductor transition temperature.

\section{Josephson current for Low-$T_c$ superconductors}
\begin{figure}[t]
\begin{center}
\includegraphics[width=8.5cm]{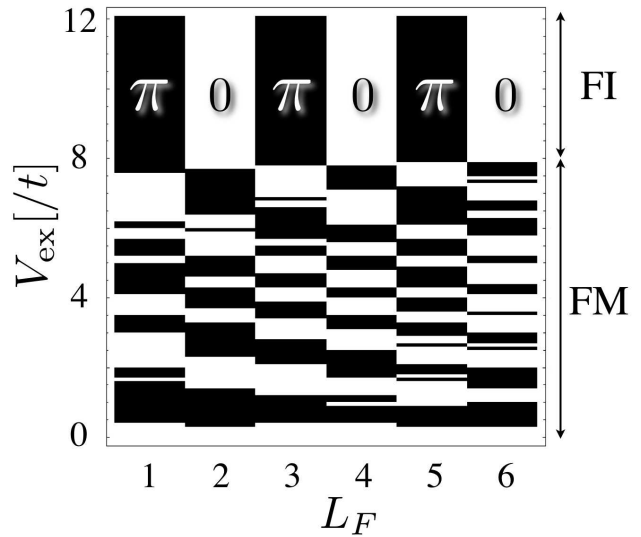}
\end{center}
\caption{The phase diagram depending on the strength of $V_\mathrm{ex}$ and $L_F$ for FM ($0 \le V_\mathrm{ex}/t  \le 8$) and FI ($V_\mathrm{ex}/t > 8$). 
The black and white regime correspond to the $\pi$- and 0-junction, respectively.}
\label{fig3}
\end{figure}

In this section we show numerical results of the Josephson current for low-$T_c$ superconductor/FI/low-$T_c$ superconductor junctions and discuss the physical origin of the $\pi$-junction formation in such systems.\cite{rf:Kawabata4,rf:Kawabata5,rf:Kawabata6}
In the calculation, we assume $t=t_s$ and set $W=25$, and $\Delta=\Delta_{s}=0.01t$. 
The phase diagram depending on the strength of $V_\mathrm{ex}$ ($0 \le V_\mathrm{ex}/t  \le 8$ for FM and $V_\mathrm{ex}/t > 8$ for FI) and $L_F$ is shown in Fig.~\ref{fig3}.
The black (white) regime corresponds to the $\pi$- (0-)junction, i.e., 
\begin{equation}
I_J=  -(+) \left| I_C \right| \sin \phi
.
\end{equation}
In the case of FI, the $\pi$-junction can be formed.
Moreover, the 0-$\pi$ transition is induced  by increasing the thickness of the FI barrier $L_F$ and the period of the transition is $universal$ and just  $single$ $atomic$ $layer$.\cite{rf:Kawabata6}
We also found that the atomic-scale 0-$\pi$ transition is also thermally stable.\cite{rf:Kawabata7}
On the her hand, in the case of FM, the oscillation period strongly depends on $V_\mathrm{ex}$ and the temperature.\cite{rf:Golubov,rf:Buzdin1}

A physical origin of the appearance of the $\pi$-junction and the atomic scale 0-$\pi$ transition can be explained as follows.\cite{rf:Kawabata6}
In the high barrier limit ($V_\mathrm{ex} \gg t$), Josephson critical current is perturbatively given by\cite{rf:Kawabata3,rf:Kawabata4} 
 \begin{eqnarray}
I_C \propto T_\downarrow^* T_\uparrow.
  \end{eqnarray}
Here $T_{\uparrow(\downarrow)}$ is a transmission coefficient of the FI barrier for up (down) spin electrons. 
In the case of the single-cite FI (i.e., $L_F=1$), the transmission coefficients are analytically given by use of the transfer matrix method\cite{rf:Sawada,rf:Xu,rf:Usuki} as 
 \begin{eqnarray}
T_{\uparrow} & =& \alpha_1 \frac{ t }{V_\mathrm{ex}},\\
T_{\downarrow} &=&   -  \alpha_1  \frac{ t }{V_\mathrm{ex}}
,
  \end{eqnarray}
where $\alpha_1$ is a spin-independent complex number.
Therefore  the sigh of the critical current
 \begin{eqnarray}
I_C \propto - | \alpha_1|^2 \left( \frac{t }{V_\mathrm{ex}} \right)^2
  \end{eqnarray}
becomes $negative$, so the $\pi$-junction is formed in the case of single-cite FI barrier.

On the other hand, the transmission coefficients for an arbitrary value of  $L_F \ge 1$ can be expressed  by  
 \begin{eqnarray}
T_{\uparrow} & =&\alpha_{L_F} \left(  \frac{ t }{V_\mathrm{ex}} \right)^{L_F},\\
T_{\downarrow} &=&  \alpha_{L_F} \left( - \frac{ t }{V_\mathrm{ex}} \right)^{L_F}
,
  \end{eqnarray}
where $\alpha_{L_F}$ is a complex number.
So the sign of the critical current 
 \begin{eqnarray}
I_C \propto  (-1)^{L_F}  | \alpha_{L_F}|^2 \left( \frac{t }{V_\mathrm{ex}} \right)^{2 L_F}
  \end{eqnarray}
becomes negative for the odd number of $L_F$ and positive for the even number of $L_F$.
Therefore we can realize the atomic-scale 0-$\pi$ transition with increasing the thickness of the FI barrier $L_F$ as demonstrated in Fig.~\ref{fig3}.

\section{Josephson current for High-$T_c$ superconductors}
\begin{figure}[b]
\begin{center}
\includegraphics[width=8.8cm]{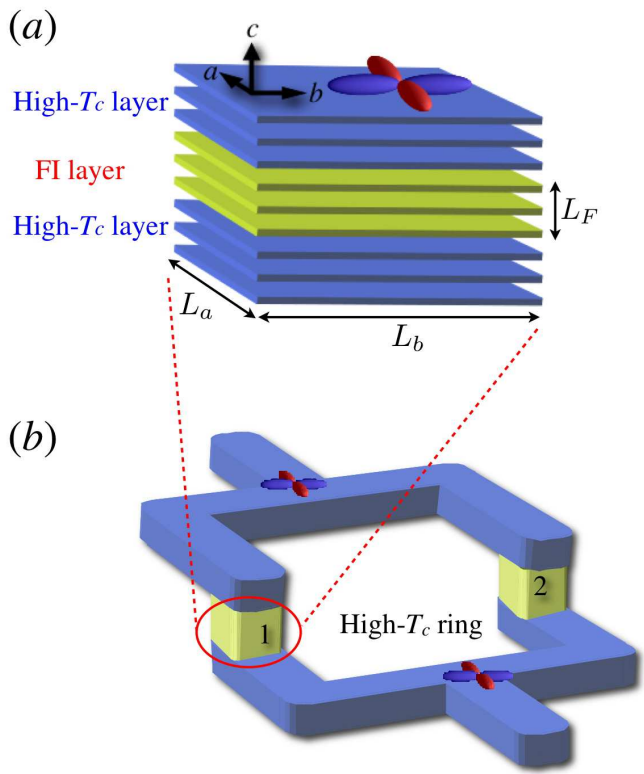}
\end{center}
\caption{Schematic picture of (a) $c$-axis stack high-$T_c$ superconductor/LBCO/high-$T_c$ superconductor Josephson junction and (b) high-$T_c$ ring which can be used in experimental observations of  the $\pi$-junction.}
\label{fig4}
\end{figure}

We would like to show an experimental set-up for observing the $\pi$-junction using LBCO in Fig.~\ref{fig4}.
From the perspectives of the FI/superconductor  interface matching  and the high-temperature device-operation, the usage of high-$T_c$ cuprate superconductors (HTSC), e.g., YBa${}_2$Cu${}_3$O${}_{7-\delta}$ and La${}_{2-x}$Sr${}_x$CuO${}_4$(LSCO) is desirable.
Recent development of the pulsed laser deposition technique enable us to layer-by-layer epitaxial-growth of such oxide materials.\cite{rf:Mercey,rf:Prellier}
Therefore, the experimental observation of the 0-$\pi$ transition by increasing the layer number of LBCO could be possible.

In order to show the possibility of $\pi$-coupling in such realistic HTSC junctions, we have numerically calculated the $c$-axis Josephson critical current $I_C$ based on a three-dimensional tight binding model with $L_a$ and $L_b$ being the numbers of lattice sites in $a$ and $b$ directions [Fig.~\ref{fig4} (a)].\cite{rf:Kawabata6,rf:Kawabata8}
In the calculation we have used a hard wall boundary condition for the $a$ and $b$ direction and taken into account the $d$-wave order-parameter symmetry in HTSC, i.e., 
 \begin{eqnarray}
\Delta= \frac{\Delta_d}{2}(\cos k_x a- \cos k_y a) .
  \end{eqnarray}
 The tight binding parameters $t$ and $g$  have been determined  by fitting to the first-principle band structure calculations~\cite{rf:LBCO}.
Figure 5 shows the FI thickness $L_F$ dependence of $I_C$ at $T=0.01 T_c$ for a LSCO/LBCO/LSCO junction with $V_\mathrm{ex}/t=28$, $\Delta_d/t=0.6$, and $L_a=L_b=100$.
As expected, the atomic scale  0-$\pi$ transitions can be realized in such oxide-based $c$-axis stack junctions.
\begin{figure}[b]
\begin{center}
\includegraphics[width=8.8cm]{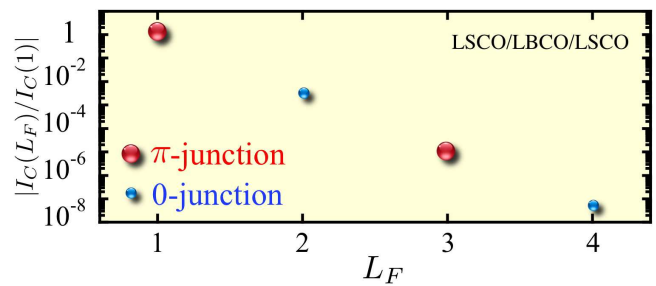}
\end{center}
\caption{The Josephson critical current $I_C$ as a function of the FI thickness $L_F$ at $T=0.01 T_c$ for a $c$-axis stack LSCO/LBCO/LSCO junction with $V_\mathrm{ex}/t=28$, $\Delta_d/t=0.6$, and $L_a=L_b=100$.
The red (blue) circles indicate the $\pi$(0)-junction.}
\label{fig5}
\end{figure}

The formation of the $\pi$-junction can be experimentally detected by using a HTSC ring [see Fig. 4 (b)].
The phase quantization condition for the HTSC ring is given by 
 \begin{eqnarray}
 2 \pi \frac {\Phi -\Phi_\mathrm{ext} }{ \Phi_0} + \phi_1 + \phi_2 = 2 \pi n
,
  \end{eqnarray}
 where
$\phi_1$ and $\phi_2$ are  the phase difference across the junction 1 and 2, $\Phi$ is the flux penetrating though the ring, $\Phi_0$ is the flux quantum,  and $n$ is an integer.
The current passed through the ring divides between the junction 1 and 2, $i.e.,$ 
 \begin{eqnarray}
  I=I_{C1} \sin \phi_1 +I_{C2} \sin \phi_2
.
  \end{eqnarray}
Applied external magnetic flux $\Phi_\mathrm{ext}$ depletes phases $ \phi_1$ and $ \phi_2$ causing interference between currents through the junctions 1 and 2.
For a symmetric ring with $I_{C1} \approx I_{C2} = I_C$ and negligible geometric inductance ($L =0$), the total critical current as a function of $\Phi_\mathrm{ext}$  is given by 
 \begin{eqnarray}
 I_C^{00} =I_C^{\pi \pi}=2 I_C \left|  \cos \left( \pi \frac{ \Phi_\mathrm{ext} }{ \Phi_\mathrm{0} } \right) \right| 
,
  \end{eqnarray}
for the case that $L_F$ of the both junctions are same.
If $L_F$ of the junction 1(2) is even and $L_F$ of the junction 2(1) is odd, we get
 \begin{eqnarray}
I_C^{0 \pi} =I_C^{\pi 0}=2 I_C \left|  \sin \left( \pi \frac{ \Phi_\mathrm{ext} }{ \Phi_\mathrm{0} } \right) \right| 
.
  \end{eqnarray}
Therefore the critical current of a 0-$\pi$ (0-0) ring has a minimum (maximum) in zero applied magnetic field.~\cite{rf:Sigrist}
Experimentally, the half-periodic shifts in the interference patterns of the HTSC ring can be used as a strong evidence of the $\pi$-junction.
Such a half flux quantum shifts have been observed in a $s$-wave ring made with a S/FM/S~\cite{rf:Guichard} and a S/quantum dot/S junction.~\cite{rf:Dam}

It is important to note  that in the case of $c$-axis stack HTSC Josephson junctions,\cite{rf:Kleiner,rf:Yurgens} no zero-energy Andreev bound-states\cite{rf:Kashiwaya} which give a strong Ohmic dissipation\cite{rf:KawabataMQT1,rf:KawabataMQT2,rf:KawabataMQT3} are formed.
Moreover, the harmful influence of nodal-quasiparticles due to the $d$-wave order-parameter symmetry on the macroscopic quantum dynamics in such $c$-axis junctions is found to be week both theoretically\cite{rf:Fominov,rf:Amin,rf:KawabataMQT4,rf:KawabataMQT5,rf:Umeki,rf:KawabataMQT5} and experimentally.\cite{rf:InomataMQT,rf:Jin,rf:Matsumoto,rf:KashiwayaMQT}
Therefore  HTSC/LBCO/HTSC $\pi$-junctions would be a good candidate for quiet qubits.

\section{Summary}
To summarize, we have studied the Josephson effect in S/FI/S junction by use of the recursive Green's function method.
We found that the $\pi$-junction and the atomic scale 0-$\pi$ transition is realized in such systems.
By use of the transfer matrix calculation, the origin of the $\pi$-junction formation can be attributed to the $\pi$ phase difference of the spin-dependent transmission coefficient for the FI barrier.
Such FI based $\pi$-junctions may become an  element in the architecture of quiet qubits.

We would like to point out that the $\pi$-junction can be also realized in the Josephson junction through an another type of FI, i.e., a spin-filter material, in the case of the strong hybridization between localized and conduction electrons.\cite{rf:Kawabata9,rf:Kawabata10}
It should be also note that FI materials treated in this paper can be categorized in strongly correlated systems.
Moreover, in actual junctions, the influence of the interface roughness could be important.
Therefore investigation of the atomic-scale 0-$\pi$ transition in the presence of the many-body and disorder effect will be also the subject of future studies.

\section*{Acknowledgements}
This paper is based on the collaboration works with S. Kashiwaya, Y. Tanaka, and A. A. Golubov.
We  would like to thank J. Arts, A. Brinkman, M. Fogelstr\"om, H. Ito, T. Kato, P. J. Kelly, T. L\"ofwander, T. Nagahama, F. Nori, J. Pfeiffer, A. S. Vaenko and M. Weides for useful discussions.
This work was  supported by CREST-JST, and a Grant-in-Aid for Scientific Research from the Ministry of Education, Science, Sports and Culture of Japan (Grant No. 22710096).

\appendix

%
%
%
%
\end{document}